\documentclass[conference]{IEEEtran}
\IEEEoverridecommandlockouts

\usepackage[T1]{fontenc}
\usepackage[utf8]{inputenc}
\usepackage{amsmath,amssymb,amsfonts}
\usepackage{graphicx}
\usepackage{cite}
\usepackage{url}
\usepackage{enumitem}

\begin{document}

\title{Towards Quantum Optimised Malware Containment}

\author{%
\IEEEauthorblockN{Matthew Sutcliffe\IEEEauthorrefmark{1}, Ravindra Mutyamsetty\IEEEauthorrefmark{1}\textsuperscript{\dag}}
\IEEEauthorblockA{\IEEEauthorrefmark{1}Intract\={a}bilis, London, UK}
}

\maketitle
\begingroup
\renewcommand\thefootnote{\dag}
\footnotetext{Corresponding Author.}
\endgroup

\begin{abstract}
The containment of malware in computing networks may be naturally formulated as a network influence minimisation problem, in which one seeks to limit the expected spread of an infection while balancing the operational cost of disabling network connections. Classical approaches often rely on Monte Carlo simulation of stochastic diffusion processes and greedy optimisation over candidate edge removals, resulting in significant computational overhead due to repeated influence evaluations. In this work, we propose a hybrid quantum approach which combines Quantum Amplitude Estimation (QAE) and Grover Minimum Finding (GMF) to provide quadratic improvements in both the estimation and optimisation components of the problem. Specifically, QAE replaces classical Monte Carlo simulation, reducing the sampling complexity of influence estimation from $O(1/\varepsilon^2)$ to $O(1/\varepsilon)$ for a target additive error $\varepsilon \ll 1$, while GMF reduces the number of candidate evaluations required to identify optimal edge removals from $O(|E_C|)$ to $O(\sqrt{|E_C|})$. We present a formal problem definition, describe the construction of the corresponding quantum oracles, and analyse the resulting complexity improvements under standard oracle assumptions. Preliminary experiments, including classical simulation of QAE and small-scale execution of Grover search on real quantum hardware, support the expected theoretical scaling. While practical implementation at scale requires fault-tolerant quantum devices, our results demonstrate that quantum algorithms offer a promising long-term direction for accelerating stochastic network optimisation problems such as malware containment.
\end{abstract}

\section{Introduction}

The spread of malware through a network is a ubiquitous cybersecurity problem, often with very damaging consequences. Once one node is compromised, an infection can often propagate throughout the network in a stochastic manner by exploiting communication channels between devices. Containing this spread is a critical challenge, requiring interventions that limit infection while preserving the operational integrity of the network.

While extreme solutions, such as shutting down and reverting the whole system, may resolve such security breaches, in many cases such drastic measures are not viable due to the importance of maintaining operations. Shutting down vital nodes, even temporarily, may cause considerable financial losses to large business networks, or even loss of life in healthcare networks. As such, a more tactical response is desirable.

In fact, this problem can be naturally formulated as a network influence minimisation task, with an objective to minimise the spread of an infection originating from a set of compromised nodes while also reducing the operational impact of shutting down important nodes. A typical classical approach to this problem may rely on greedy iteration of a stochastic diffusion model, such as the Independent Cascade model, combined with Monte Carlo simulation to estimate expected influence.

This framing is useful because it isolates the central trade-off in a mathematically explicit way. On the one hand, defenders wish to reduce the expected number of compromised machines as quickly as possible. On the other hand, defensive interventions themselves incur cost: disabling a high-value communication link may prevent further spread, but may simultaneously disrupt production systems, clinical workflows, or time-critical business processes. Casting the problem as an optimisation over a graph therefore provides a natural language for balancing security benefit against operational disruption.

In this paper, we outline how such an approach can be improved with quantum computation, allowing, in theory, for quadratically better scaling in both solution accuracy and candidate search space. In particular, we employ a combination of Quantum Amplitude Estimation (QAE) and Grover Minimum Finding (GMF) to this end, and --- while practical application remains infeasible until fault tolerant quantum computers become available --- we nevertheless demonstrate the potential viability of our approach in small-scale prototype experiments on real quantum hardware.

Conceptually, the proposed hybrid quantum workflow separates into two coupled subproblems. The first is an estimation problem: for any proposed intervention, one must estimate the residual influence of the infected seed set under a stochastic diffusion process. The second is an optimisation problem: among many possible interventions, one must identify the one that most improves the objective. Classically, both steps are expensive because each candidate action typically requires a fresh stochastic estimate. The appeal of the quantum setting is that QAE targets the estimation bottleneck, while GMF targets the search bottleneck.

\section{Problem Formulation and Quantum Preliminaries}

While there are innumerable ways to formulate a network influence minimisation problem --- and a malware containment problem more broadly --- we define a version as follows.

\subsection{Problem Formulation}

Given a graph $G = (V, E)$ with:
\begin{itemize}[leftmargin=*,nosep]
    \item edge activation probabilities $p : E \to [0,1]$,
    \item a seed set of nodes $S \subseteq V$, and
    \item edge operational importance $i : E \to [0,1]$,
\end{itemize}
find a set of edges $E' \subseteq E$ which minimises the objective function:
\[
\lambda\sigma(S; G\setminus E', p\setminus E') + (1-\lambda)OI(S; E', i)
\]
where:
\begin{itemize}[leftmargin=*,nosep]
    \item $\sigma(S; G\setminus E', p\setminus E')$ is the expected influence of $S$ after removing the edges $E'$ from $G$, with $G\setminus E' := (V, E\setminus E')$ and $p\setminus E' := p(e)\ \forall e \in E\setminus E'$,
    \item $OI(E', i) := \sum_{e\in E'} i_e$ is the operational impact of removing the edges $E'$, and
    \item $\lambda$ denotes a fixed weighting coefficient in the range $[0,1]$ indicating the balance of importance between $\sigma$ and $OI$.
\end{itemize}

\begin{figure}[t]
    \centering
    \includegraphics[width=0.78\columnwidth]{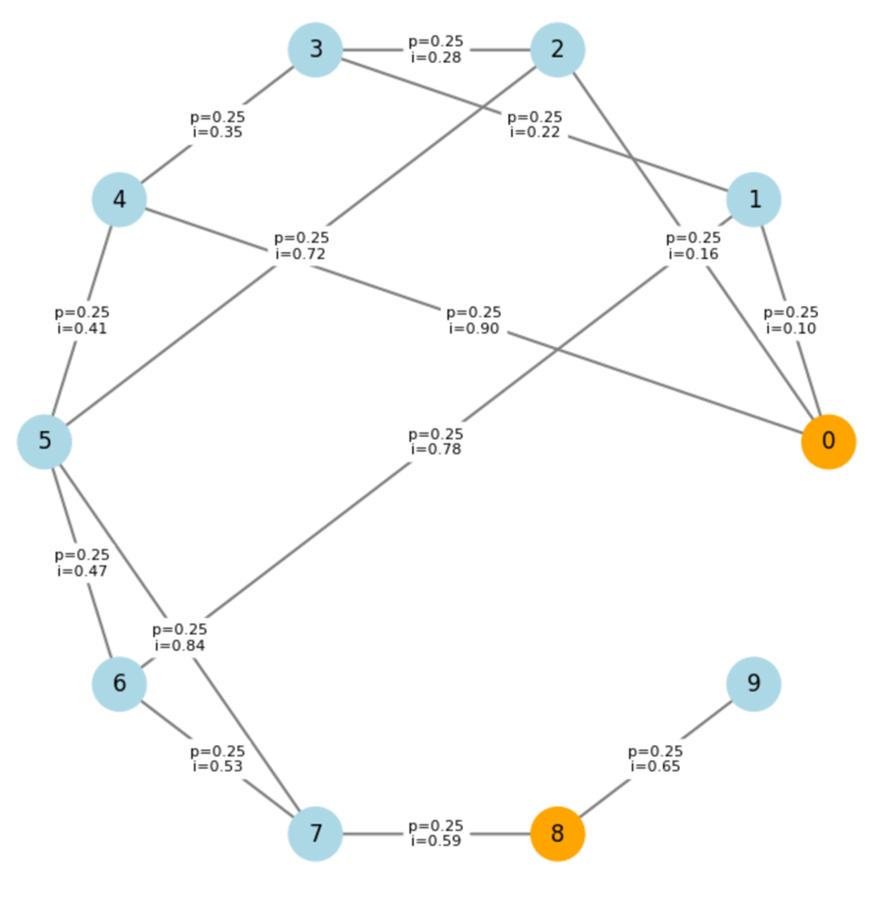}
    \caption{An example of a graph with initial infected seed nodes highlighted in orange, and the activation probabilities and operational importances of each edge indicated.}
    \label{fig:graph}
\end{figure}

\noindent\rule{\columnwidth}{0.4pt}
\par Translating this formulation to the real world, the graph represents a network of computers or other devices (nodes) with some pairwise data connections (edges) between them. Each such connection has an associated operational importance $i$ which quantifies how vital its continued operation is to the system, such that a near-zero value denotes a connection which may freely be shut down with little consequence and, conversely, a value near $1$ denotes an essential connection whose disabling would cause great cost (financial or otherwise) to the organisation. The seed set $S$ then denotes a subset of these devices on which malware has been detected, and each communication channel (edge) between a pair of devices has an associated activation probability quantifying how secure ($p \approx 0$) or vulnerable ($p \approx 1$) this connection is and determining the probability, at each time step, that the malware may spread between these devices. As such, the problem becomes one of determining which connections to shut down to minimise the spread of malware while also minimising the operational impact of doing so.

The weighting parameter $\lambda$ has an especially important interpretive role. When $\lambda$ is close to $1$, the optimisation prioritises limiting malware spread even at substantial operational cost. When $\lambda$ is close to $0$, the formulation instead becomes conservative with respect to network disruption, favouring only low-cost interventions. In practice, this parameter would be chosen according to organisational risk appetite, criticality of the affected systems, and the stage of the incident response process.

\subsection{Classical Greedy Baseline}

While the operational impact term is trivial to compute, determining the expected influence of the graph is less straightforward. A common approach is to average over a Monte Carlo (MC) simulation with an Independent Cascade model. Specifically, the Independent Cascade model, $IC(G, p, S)$, is a stochastic process defined as follows:
\begin{itemize}[leftmargin=*,nosep]
    \item Initialisation: At time step $t=0$, each seed node $v_s\in S$ is activated and each non-seed node remains inactive.
    \item Diffusion steps:
    \begin{itemize}[leftmargin=1.2em,nosep]
        \item At each step $t\geq 1$, each node $v$ that is activated in the previous step $t-1$ activates each of its inactive neighbours $u$ with activation probability $p(v,u)$.
        \item That is, each activated node remains active for the whole process but can only activate other nodes one step after its activation.
        \item The process terminates when no node is activated in the previous step.
    \end{itemize}
\end{itemize}

With this, the influenced probability of a vertex $v$, denoted $\pi(v; G, p, S)$, is the probability of $v$ being influenced (i.e. active) when the process of $IC(G, p, S)$ terminates. Thus, the expected influence of a seed set $S$, denoted by $\sigma(S; G, p)$ is defined as the expected number of finally influenced nodes,
\[
\sigma(S; G, p) := \sum_{v\in V} \pi(v; G, p, S).
\]

As such, a simple greedy classical solution to this overall network influence minimisation problem may proceed as follows:
\begin{itemize}[leftmargin=*,nosep]
    \item Determine, via some heuristic strategy, a subset of `candidate' edges $E_C \subseteq E$ to consider for removal.
    \item For each candidate edge ($\forall e\in E_C$), consider the corresponding graph with this edge removed, $G\setminus e := (V, E\setminus e)$, and estimate its expected influence, $\sigma(S; G\setminus e, p\setminus e)$ via IC/MC.
    \item Select the candidate edge whose removal minimises the expected influence, i.e.
    select $e$ for which $\min_{e\in E_C} \sigma(S; G\setminus e, p\setminus e)$, and remove this edge from the graph, $G \rightarrow G\setminus e$.
    \item Repeat this process until no suitable candidate edges are found or until further edge removals fail to reduce the expected influence.
\end{itemize}

The practical difficulty in this baseline lies in the repeated influence estimation. Even when only a modest candidate set is considered, each greedy step may require many Monte Carlo evaluations, and each evaluation itself simulates a random diffusion process over the graph. Consequently, the total cost compounds across candidate edges, across repeated Monte Carlo samples, and across greedy iterations. This layered cost structure is exactly what makes the problem a natural target for algorithms that improve either estimation complexity or search complexity.

\subsection{Quantum Computing Background}

An $n$-qubit quantum state, $|\psi\rangle$, may be described by a normalised vector of $2^n$ complex amplitudes:
\[
|\psi\rangle = \sum_{x\in\{0,1\}^n} \alpha_x |x\rangle,
\]
where $\alpha_x \in \mathbb{C}\ \forall x$ and $\sum_x |\alpha_x|^2 = 1$. If more than one computational basis state, $|x\rangle$, has a non-zero amplitude, $\alpha_x$, then the quantum state is said to be in a superposition, being essentially in multiple classical configurations simultaneously~\cite{nielsen2010}.

Encoding information with qubits rather than classical bits therefore allows for the representation of superpositions of exponentially many classical configurations, enabling quantum algorithms to exploit interference effects that have no classical analogue.

For the purposes of this paper, the importance of the quantum formalism is not merely representational but algorithmic. Superposition allows many computational branches to be prepared coherently at once, while interference allows those branches to be combined so that desirable outcomes are amplified and undesirable ones are suppressed. The speedups discussed later arise not from evaluating every possibility independently in parallel, but from designing unitary transformations whose global interference pattern reveals information about the solution more efficiently than repeated classical sampling.

Many such quantum algorithms are formulated in the oracle model, where an oracle is a black-box unitary operation that encodes a function $f : \{0,1\}^n \to \{0,1\}$. A typical implementation applies a phase flip to states $|x\rangle$ for which $f(x)=1$:
\[
O_f|x\rangle = (-1)^{f(x)}|x\rangle.
\]

\subsection{Grover Search}

The well-known Grover's algorithm~\cite{grover1996} is a staple of quantum computing, providing a quadratic speedup versus classical methods to the problem of searching an unsorted list. Specifically, in searching for a particular element within an unsorted list of length $N$, any classical algorithm must inherently require an $O(N)$ linear search through the list. By contrast, in making use of superposition and interference (properties unique to quantum mechanics) the quantum Grover's algorithm is able to find the element in $O(\sqrt{N})$ with a high constant probability (which may be increased arbitrarily close to unity through repetition).

The key idea underpinning Grover's algorithm is amplitude amplification. Starting with a uniform superposition:
\[
|s\rangle = \frac{1}{\sqrt{N}} \sum_x |x\rangle,
\]
it aims to amplify the amplitude of the `marked' states (those for which $f(x)=1$). It does this through repeated iterations of:
\[
G = DO_f,
\]
where $O_f$ is the oracle and:
\[
D = 2|s\rangle\langle s| - I
\]
is the diffusion operator (with $I$ denoting the identity operator). Repeated applications of $G$ thus redistributes the amplitude via interference, with marked states interfering constructively and unmarked states interfering destructively. Finally, after $O(N)$ iterations, measuring the state yields a marked element with high probability.

At a high level, Grover's search can therefore be understood as replacing exhaustive trial-and-check over an unstructured set with a sequence of coherent rotations in a low-dimensional subspace spanned by marked and unmarked states. This geometric viewpoint is useful because it clarifies why the improvement is quadratic rather than exponential: the algorithm does not remove the need to query the oracle, but it reduces the number of such queries required to concentrate probability mass on the desired answers.

\subsection{Grover Minimum Finding}

Grover's algorithm can be extended to optimisation problems, such as finding the minimum (or maximum) element of an unsorted list. This variant, named for its authors, is known as the D"urr-H{
m	extbackslash o}yer algorithm~\cite{durr1996}, or commonly Grover minimum/maximum finding (GMF). In particular, given a function $g : \{1,\ldots,N\} \to \mathbb{R}$, it aims to find:
\[
\min_{i\in\{1,\ldots,N\}} g(i)
\]
using $O(\sqrt{N})$ evaluations of $g$, as compared to the $O(N)$ evaluations required of classical algorithms.

In brief, the algorithm works by maintaining a current candidate minimum value $g(i^\star)$ and iteratively improving it by repeating the following steps:
\begin{itemize}[leftmargin=*,nosep]
    \item define an oracle that marks all indices $i$ such that $g(i) < g(i^\star)$,
    \item use Grover search to find such an index (if one exists), and
    \item update $i^\star$ if a better candidate is found.
\end{itemize}
Thus, each Grover search requires $O(\sqrt{N})$ oracle calls and the overall algorithm succeeds with high probability using $O(\sqrt{N})$ evaluations of $g$.

For malware containment, the role of the function $g$ is played by an objective derived from the residual influence after applying a candidate intervention. The optimisation problem is therefore not over arbitrary numbers, but over edge removals whose quality depends on a stochastic process on the underlying graph. This is important because it means the GMF oracle must in effect compare interventions according to estimated influence, making the quality of the estimation subroutine central to the success of the overall method.

\subsection{Quantum Amplitude Estimation}

In a similar vein, there exist a family of quantum algorithms that can provide quadratic speedups for estimating the expectations of random variables through Quantum Amplitude Estimation (QAE)~\cite{brassard2000}.

Simply put, suppose one wished to estimate the expectation of a bounded random variable $X\in[0,1]$. Classically, this would typically be computed by averaging over repeated sampling:
\[
\mathbb{E}[X] \approx \frac{1}{T}\sum_{i=1}^{T} X_i,
\]
which requires $T = O(1/\varepsilon^2)$ samples to achieve an additive error $\varepsilon$. By contrast, QAE is able to achieve the same task through only $O(1/\varepsilon)$ queries.

QAE requires a unitary operator $A$ that prepares a quantum state:
\[
A|0\rangle = \sqrt{1-a}|\psi_0\rangle|0\rangle + \sqrt{a}|\psi_1\rangle|1\rangle,
\]
where $a\in[0,1]$ is an unknown value to be estimated and the final qubit indicates `success' ($|1\rangle$) or `failure' ($|0\rangle$).

It requires also an operator for reflection about the initial state:
\[
S_0 = I - 2|0\rangle\langle 0|,
\]
and one for reflection about the `good' subspace (where the last qubit is $|1\rangle$):
\[
S_f = I - 2(I\otimes |1\rangle\langle 1|),
\]
such that the amplitude amplification operator may then be given by:
\[
Q = AS_0A^{\dagger}S_f.
\]

Given a quantum state:
\[
A|0\rangle = \sin(\theta)|\psi_1\rangle|1\rangle + \cos(\theta)|\psi_0\rangle|0\rangle,
\]
where $a = \sin^2(\theta)$, each application of $Q$ rotates the state by an angle $2\theta$ such that after $k$ applications:
\[
Q^kA|0\rangle = \sin((2k+1)\theta)|\psi_1\rangle|1\rangle + \cos((2k+1)\theta)|\psi_0\rangle|0\rangle.
\]
The goal of QAE is to estimate $\theta$ and hence $a = \sin^2(\theta)$. This is typically achieved with quantum phase estimation applied to the operator $Q$. Since $Q$ has eigenvalues $e^{\pm 2i\theta}$, phase estimation allows $\theta$ to be estimated to additive error $\varepsilon$ using $O(1/\varepsilon)$ applications of $Q$. Thus, QAE achieves a query complexity of $O(1/\varepsilon)$, compared to the classical Monte Carlo complexity of $O(1/\varepsilon^2)$.

QAE can be applied to many expectation estimation problems, provided they can be reduced to amplitude estimation problems. In particular, given a random variable $X\in[0,1]$, one can construct a unitary $A$ such that:
\[
\mathbb{E}[X] = a,
\]
where $a$ is the amplitude of measuring $|1\rangle$ in the final qubit. More specifically, this involves a superposition over random samples and encodes the value of $X$ into a qubit rotation. Measuring this qubit yields $|1\rangle$ with probability equal to $\mathbb{E}[X]$.

This observation makes QAE especially relevant for stochastic network processes. If the randomness of the diffusion model can be encoded coherently, then the expected spread of an infection may be treated as an amplitude to be estimated rather than as a quantity that must be approximated by many independent classical trials. In that sense, QAE is not changing the underlying influence model; it is changing the computational mechanism by which the expectation associated with that model is obtained.

\section{Quantum Approach}

Returning to the network influence minimisation problem outlined in section II-A, we propose applying a combination of Grover Minimum Finding (section II-E) and Quantum Amplitude Estimation (section II-F) to provide quadratic improvements to the computational complexity with respect to both the graph size and the error rate.

\subsection{Influence Estimation via QAE}

In the classical setting, influence is typically estimated by averaging over repeated stochastic simulations of the diffusion process. For a target additive error $\varepsilon \ll 1$ (i.e.
a fraction of the total number of graph nodes), this requires $O(1/\varepsilon^2)$ samples.

Instead, we propose employing QAE here in place of the IC/MC method as a means of estimating the influence of the graph, which can provide a quadratic improvement, reducing this scaling to just $O(1/\varepsilon)$ oracle calls.

The approach proceeds as follows:
\begin{itemize}[leftmargin=*,nosep]
    \item Construct a unitary operator $A$ that prepares a superposition over the randomness of the diffusion process (or a suitable surrogate such as a live-edge realisation).
    \item Encode the outcome of interest (e.g.
    number or fraction of infected nodes) into a designated ancillary qubit, such that the probability of measuring $|1\rangle$ is equal to the expected influence.
    \item Apply QAE to estimate this probability using repeated controlled applications of $A$ and $A^{\dagger}$.
\end{itemize}

This yields an estimate of the expected influence with additive error $\varepsilon$ using $O(1/\varepsilon)$ oracle calls, compared to $O(1/\varepsilon^2)$ classical samples of IC/MC simulation.

Thus, we target the dominant computational bottleneck and achieve a means of estimating the influence with quadratically fewer samples for the same accuracy. In addition to this benefit, computing these influences within a superposition in an oracle is, as will be shown, vital for enabling the next quantum method we propose.

From a modelling standpoint, the central technical requirement is therefore the construction of an operator $A$ that faithfully represents the diffusion randomness and the statistic of interest. In a full implementation, this would typically require reversible encodings of graph structure, edge activation randomness, propagation logic, and a normalised influence score. The present paper focuses on the complexity-theoretic consequences of such an oracle construction rather than on a hardware-efficient circuit synthesis, but this distinction is important: the asymptotic speedup assumes access to a suitable oracle, while the practical cost depends heavily on how expensive that oracle is to realise.

\subsection{Greedy Edge Selection via GMF}

In the classical baseline, after the influence has been estimated for every potential edge removal amongst the candidate set (i.e.
$\sigma(S; G\setminus e, p\setminus e)\ \forall e\in E_C$), the edge whose removal minimises the influence the most can be determined through a simple $O(|E_C|)$ linear search. Alternatively, Quantum Minimum Finding (QMF) could identify the optimal candidate using only $O(\sqrt{|E_C|})$ oracle calls.

In particular, the workflow could proceed as follows:
\begin{itemize}[leftmargin=*,nosep]
    \item Define a candidate set of edges $E_C \subseteq E$.
    \item Construct an oracle that, given an edge $e\in E_C$, estimates the marginal influence of the corresponding graph with this edge removed and marks it if it exceeds a given threshold.
    \item Apply Grover-style amplitude amplification to identify candidates exceeding the threshold.
    \item Iteratively update the threshold to converge to the optimal edge.
\end{itemize}

This yields a quadratic reduction in the number of objective evaluations from $O(|E_C|)$ to $O(\sqrt{|E_C|})$ in determining the (greedily) optimal edge to remove at each step.

Importantly, to achieve this benefit, one requires an oracle that is able to estimate these network influences in a superposition. If instead the influences are estimated classically (such as via IC/MC) then the advantage of an $O(\sqrt{|E_C|})$ search through the $|E_C|$ candidates is lost as computing each would still require $O(|E_C|)$ computations beforehand). Fortunately, such an oracle can be provided by QAE as described in section III-A.

This dependence between the two quantum ingredients is worth emphasising. GMF by itself accelerates the search over candidate interventions only when the value associated with each candidate can itself be queried coherently. QAE provides precisely the mechanism for promoting a classically estimated expectation into a quantum-accessible quantity. The contribution of the combined approach is therefore not merely that two well-known quantum algorithms are used side by side, but that they are composed in a way that addresses both nested levels of the classical workload.

\section{Results and Discussion}

\subsection{Theoretical Complexity Analysis}

With the use of QAE and GMF as described in this paper, the network influence minimisation problem detailed in section II-A can --- in theory --- be solved with quadratically better scaling against both error rate and candidate edge count. Specifically, the theory suggests a complexity improvement from a classical scaling of:
\[
T_{\mathrm{classical}} = O\!\left(k\cdot |E_C| \cdot T_{\mathrm{diff}} \cdot \frac{1}{\varepsilon^2}\right)
\]
to a quantum-enabled scaling of:
\[
T_{\mathrm{quantum}} = O\!\left(k\cdot \sqrt{|E_C|} \cdot T_{\mathrm{oracle}} \cdot \frac{1}{\varepsilon}\right),
\]
where:
\begin{itemize}[leftmargin=*,nosep]
    \item $E_C \in E$ denotes the set of candidate edges considered for removal, as determined by any suitable heuristic method. Hence, $|E_C| \leq |E|$ denotes the number of such candidate edges.
    \item $k$ denotes the number of greedy iterations, i.e.
    the number of edges removed sequentially.
    \item $\varepsilon$ denotes the target additive error in the normalised influence, defined as the fraction of vertices which become infected. This corresponds to an absolute error of $\varepsilon_{\mathrm{nodes}} = \varepsilon |V|$ nodes.
    \item $T_{\mathrm{diff}}$ denotes the cost of a single classical influence evaluation (i.e.
    one Monte Carlo simulation of the diffusion process). In the IC/MC setting, this corresponds to a graph traversal (such as via breadth-first search) and satisfies $T_{\mathrm{diff}} = O(|V|+|E|)$.
    \item $T_{\mathrm{oracle}}$ denotes the cost of a single full quantum oracle evaluation of the influence.
\end{itemize}

In practice, the quantum oracle will be computationally expensive, $T_{\mathrm{oracle}} \gg T_{\mathrm{diff}}$, and susceptible to noise, suggesting any practical improvement will require large scale networks as well as fault tolerant hardware.

\subsection{Evaluation of QAE}

\begin{figure}[t]
    \centering
    \includegraphics[width=\columnwidth]{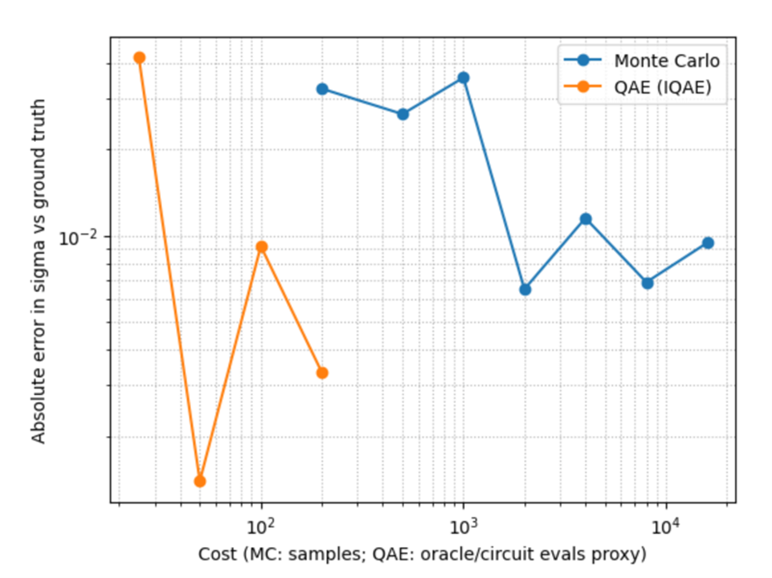}
    \caption{The accuracy $\varepsilon$ of influence estimation achieved after (a) a number of Monte Carlo runs (with the classical IC/MC method) vs (b) a number of oracle calls (with the QAE method).}
    \label{fig:qae}
\end{figure}

We prototyped the QAE implementation and tested its viability on a small randomly generated networks, using classical simulation of the quantum oracle (as appropriate quantum hardware would require fault tolerance, which is not yet available). Figure~\ref{fig:qae} shows the results for a particular randomly selected $20$ node example graph, demonstrating that the QAE solution achieves better accuracy solutions with significantly fewer iterations than the classical IC/MC approach. (In the QAE case, the number of iterations refers to the number of oracle calls, whereas in the IC/MC case it refers to the number of Monte Carlo runs.)

As shown, even with $16{,}000$ Monte Carlo iterations, the resulting accuracy (marginally) fails to reach the target $99.5\%$ accuracy (reaching only $99.31\%$), whereas QAE manages to reach this target (in fact reaching $99.86\%$) in only $50$ oracle calls. This represents a reduction in the number of iterations by a factor of $16{,}000/50 = 320$, which is an improvement more-or-less consistent with what one would expect from the theory outlined above. That is, given an accuracy target of $99.5\%$ (i.e.
$\varepsilon = 0.005$), one would expect approximately $O(1/\varepsilon^2) = O(1/0.005^2) = O(40{,}000)$ iterations of Monte Carlo or approximately $O(1/\varepsilon) = O(1/0.005) = O(200)$ iterations of QAE oracle calls, predicting an improvement factor of $40{,}000/200 = 200$, similar to the $320$ observed in the experiment.

For more context on this experiment, note that the `true' expected influence value (in this example case an average spread of $2.27250$ nodes), against which all other measured values were compared to deduce their accuracy, was determined via $200{,}000$ iterations of Monte Carlo. Also note that, as the QAE is simulated on classical hardware, it is not possible to meaningfully measure and compare the runtimes of the two methods, only their respective number of iterations.

As QAE is highly sensitive to noise, its use as a practical solution on larger, more meaningful experimental scales is infeasible without fault tolerant hardware, making this approach unsuitable as a near-term solution but potentially viable as a long-term solution. Additionally, while QAE reduces the number of samples required, each one becomes a deep coherent computation. As such, whether a meaningful runtime advantage is observed will depend upon the experimental scale and how much more expensive each sample is to compute versus classical simulation.

\subsection{Evaluation of GMF}

Figure 3 shows the results of experiments measuring the number of steps taken to find the minimum of an unsorted list of pre-computed influences for each candidate edge, using both a classical linear search and a Grover search. We used pre-computed influences here to clearly demarcate the QAE experiments from the Grover experiments and, in this case, focus on the latter. Moreover, note that the `number of steps' refers to the number of oracle calls in the case of the Grover results and simply the number of candidate edges $|E_C|$ in the case of the classical linear search.

\begin{figure}[t]
    \centering
    \includegraphics[width=\columnwidth]{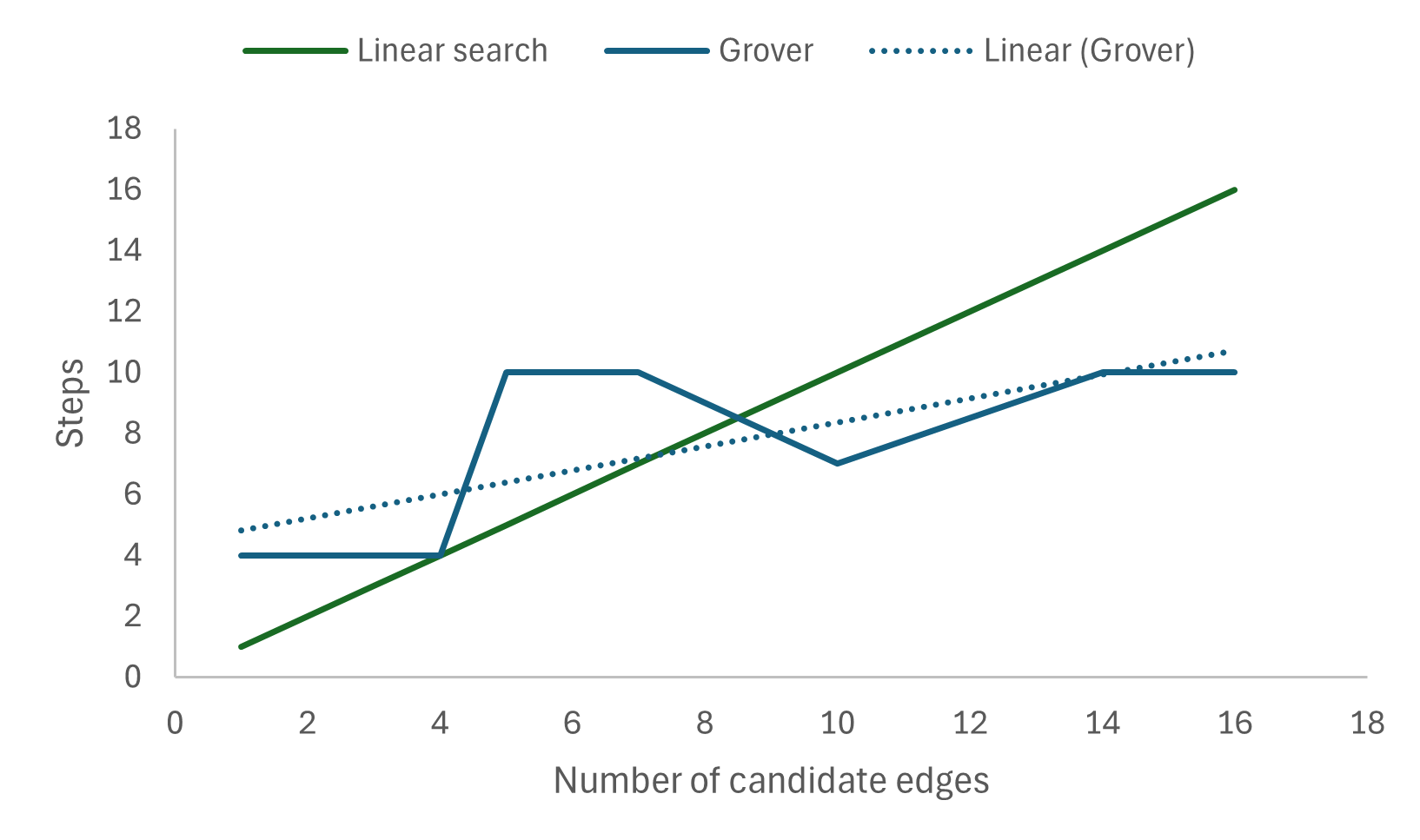}
    \caption{The number of steps (classical linear search steps or quantum oracle calls) required to find the minimum of an unsorted list of candidate edge influences. For the Grover approach, a line of best fit is included.}
    \label{fig:gmf}
\end{figure}

These Grover measurements were computed on real quantum hardware (namely IBM's `Fez') and were hence limited to small scales, involving randomly generated graphs of $2$--$10$ nodes and $1$--$16$ candidate edges. Nevertheless, even at such scales, these measurements demonstrate a clear reduction in the number of oracle calls required of the Grover approach as compared to the number of steps involved in the linear search, with a very promising trend suggesting more favourable complexity scaling.

However, GMF too should be considered as a potential long-term solution as it faces the same limitations as discussed of QAE. In particular, while GMF reduces the number of oracle calls (as compared to number of linear search steps), each such call is, in practice, expensive, meaning the improvement to the number of steps involved as shown in the figure does not indicate a corresponding improvement to runtime unless much greater experimental scales are considered. Moreover, GMF requires coherent oracle calls and precise phase rotations, making it highly sensitive to noise and therefore impractical on near-term hardware for experiments beyond very small scales. Nevertheless, future fault tolerant hardware may address these limitations by minimising noise and rendering larger scale experiments practical.

\section{Conclusion and Outlook}

This paper presents how quantum algorithms may be applied to network influence minimisation problems and, by extension, malware containment problems to achieve quadratically improved scaling in solution accuracy and in selecting greedily optimal choices among a candidate search space.

In particular, we utilised Quantum Amplitude Estimation and Grover Minimum Finding to respectively optimise the influence estimation and search over candidate edge removals, achieving the theoretical improvements highlighted in section IV-A.

We further prototyped these ideas and measured their performance on real quantum hardware at small scales, showing results largely consistent with expectation. However, these results should be treated as a preliminary proof of concept rather than indicative of ready and practical solution. There are a number of caveats to consider, most notably the limitations of present day `NISQ' hardware. Due to the noise inherent in such hardware, experiments such as ours are unable to scale beyond the very small cases demonstrated in this paper without the results becoming dominated by noise. Nevertheless, with ongoing advancements in fault tolerant architecture~\cite{bluvstein2026,ostmann2025} and protocols~\cite{gottesman2022,roffe2019}, the scalability of our methods may become practical on future quantum computers. As such, we present this work as a proof of concept of a potential long-term quantum solution.

\end{document}